\documentclass[pre,floatfix]{revtex4-2}
\pdfoutput=1 
\usepackage{graphicx}
\usepackage{latexsym}
\usepackage{amsmath, amsthm, amssymb, wasysym}
\usepackage{xcolor}
\usepackage{epsfig}
\usepackage{soul}
\usepackage{cancel}
\usepackage{hyperref}
\usepackage{orcidlink}
\usepackage{pagecolor}
\usepackage{etoolbox}

\hypersetup{
 colorlinks=true,
 linkcolor=black,
 citecolor=blue,
 urlcolor=black
}

\begin{document}
\title{A Gaussian-Remainder Hierarchy for Sums of Random Variables with Big-Jump Statistics
}

\author{Stanislav Burov\orcidlink{0000-0003-1065-174X}}

\email{stasbur@gmail.com}
\affiliation{Department of Physics, Bar-Ilan University, Ramat-Gan 5290002,
Israel}

\begin{abstract}
We develop an exact Gaussian-remainder hierarchy for the probability density of the sum of $N$ independent, identically distributed random variables with broad, finite-variance distribution for the summands. 
The hierarchy separates the Gaussian fixed-point contribution from residual sectors that retain the original single-summand density. 
For subexponential densities, the first nontrivial truncation yields a simple finite-$N$ approximation that involves one convolution with a Gaussian background and a subtraction that removes Gaussian overcounting. 
This approximation captures the Gaussian center, the crossover region, and the big-jump tail within a single expression, as demonstrated numerically for stretched-exponential and finite-variance power-law examples.
The same first-order approximation reproduces the known asymptotic anomalous rate function for sums of stretched-exponential random variables and also provides an accurate approximation to the corresponding finite-\(N\) rate function.
\end{abstract}

\maketitle

\newcommand{\fr}{\frac}
\newcommand{\lr}{\langle}
\newcommand{\rl}{\rangle}
\newcommand{\tl}{\tilde}

\section{Introduction}
\label{sec:introduction}

Approximating the probability density of a sum of $N$ independent random variables is a basic problem in statistical physics and probability theory. Such sums appear whenever an observable is built from many contributions: steps of a random walk, increments of a stochastic process, renewal events, local weights in a many-body system, or repeated contributions accumulated along a trajectory. The question addressed in this paper is how to approximate the full probability density of the sum, including both its central region and its tails.

When the single-summand distribution has a finite second moment, the central limit theorem (CLT) gives the leading description of the center of the sum distribution. In this regime the sum is described by a Gaussian, and systematic corrections around the Gaussian core are provided by Edgeworth-type expansions \cite{Edgeworth1905,Petrov1975,Hall1992}. These approximations are designed for typical fluctuations and their nearby corrections. They are not intended to describe the far tails of the sum distribution, where rare events may be produced by mechanisms that are qualitatively different from Gaussian fluctuations.
The tail behavior is especially interesting when the single-summand distribution has a broad but finite-variance tail, including subexponential laws and power-law tails with sufficiently high exponent. In such cases, the CLT still controls the center of the sum distribution, while the far tail may be governed by the big-jump principle (BJP): a large value of the sum is produced mainly by one unusually large summand rather than by the collective effect of many moderate summands \cite{Chistyakov1964,Denisov2008,Foss2013,Vezzani2019SingleBigJump}. This mechanism has attracted renewed interest because, for subexponential distributions, it is closely related to condensation-like dynamical phase transitions, where the crossover from collective fluctuations to single-event domination becomes a physical transition rather than only a question of tail estimation \cite{Gradenigo2019RunTumble,Mori2021LateTimeRTP,Mori2021RTPCondensation,Brosset2020Semiexponential,Nickelsen2018AnomalousLDP,Smith2022OU,SmithMajumdar2022Resetting,singh2023universal,shafir2024disorder}.

This coexistence of two mechanisms makes it difficult for a single standard approximation to describe both the center and the tail. The Gaussian approximation describes the center but fails in the tail, while the BJP describes the asymptotic far tail but not the full crossover back to typical fluctuations. Recent works have addressed this problem from complementary asymptotic directions. In Ref.~\cite{Smith2025Subleading}, a subleading large-$N$ theory was developed for the anomalous intermediate large-deviation regime of sums with stretched-exponential single-summand distributions, including the homogeneous, condensed, and transition regions. In Ref.~\cite{BassanoniHamdi2026BeyondBigJump}, a perturbative expansion was developed around the big-jump configuration, yielding systematic large-$X$ corrections to the BJP and extending the tail-side description toward moderate deviations.
Here we take a different route. Our goal is to construct a representation of the full finite-$N$ probability density of the sum itself, rather than an expansion only around the Gaussian center, the anomalous large-deviation regime, or the far-tail big-jump limit. We build an exact Gaussian-remainder hierarchy for the sum distribution. The hierarchy itself applies generally to finite-variance summand distributions. In the present work we focus on subexponential examples, where the competition between the Gaussian center and the big-jump tail is tied to an anomalous large-deviation structure.

The central observation of the paper is that the first nontrivial truncation of the hierarchy already gives a closed and accurate approximation. It keeps the Gaussian center and the single-large-summand mechanism in the same expression, with the subtraction needed to avoid Gaussian overcounting. We show numerically that this approximation works well across the center, the crossover region, and the tail.
The same representation also clarifies the large-deviation structure for subexponential single-summand distributions. On the anomalous large-deviation scale relevant to this class, the first-order sector of the hierarchy already contains the leading rate function and the associated condensation transition. Higher-order sectors provide finite-$N$ corrections and refinements of the probability density, but they do not change the leading rate-function structure. Thus, the hierarchy connects two aspects of the problem: (I) an accurate finite-$N$ approximation of the sum distribution and (II) the asymptotic phase-transition picture.

The paper is organized as follows. 
In Sec.~\ref{sec:hierarchy}, we derive the exact Gaussian-remainder hierarchy for the probability density of a sum of independent random variables. 
We discuss the first-order truncation and its interpretation as a Gaussian background with one explicit broad-density contribution. 
In Sec.~\ref{sec:exampleStretched}, we test the approximation for a stretched-exponential density and compare it with the exact convolution and with standard limiting descriptions. 
In Sec.~\ref{sec:anomalous_ldf}, we analyze the anomalous large-deviation scaling and show that the limiting rate function and the condensation transition already emerge from the first-order sector of the hierarchy. 
We conclude in Sec.~\ref{sec:discussion} with a summary and discussion of the relation to classical large-deviation results, the role of finite-\(N\) corrections, and possible extensions.


\section{Gaussian-Remainder Hierarchy}
\label{sec:hierarchy}
We consider the sum
\begin{equation}
    X=x_1+\cdots+x_N ,
    \label{eq:sum_def}
\end{equation}
where the $x_i$ are independent and identically distributed random variables drawn from a density $p(x)$ with mean $\mu$ and finite variance $\sigma^2$. We denote by $P_N(X)$ the probability density of the sum. Its Fourier representation is
  \begin{equation}
    P_N(X)
    =
    \int_{-\infty}^{\infty}
    \frac{dk}{2\pi}
    e^{-ikX}
    [\tilde p(k)]^N , 
    \label{eq:PN_fourier}
  \end{equation}  
    while 
    $    \tilde p(k)
    =
    \int_{-\infty}^{\infty}
    e^{ikx}p(x)\,dx $. 
Expressing $\tilde p(k)$ as $\tilde p(k)=\exp[-\omega(k)]$, the PDF of $X$ becomes
    $P_N(X)
    =
    \int_{-\infty}^{\infty}
    \frac{dk}{2\pi}
    \exp[-ikX-N\omega(k)]$ .
The Gaussian center is obtained from the small-$k$ expansion of $\omega(k)$. Since the first two cumulants of $p(x)$ are $\mu$ and $\sigma^2$, the quadratic part is
    $\omega_G(k)
    =
    -i\mu k+\frac{\sigma^2 k^2}{2}$.
We introduce the Gaussian reference density $g(x)$ through its Fourier transform,
\begin{equation}
    \tilde g(k)
    =
    \exp[-\omega_G(k)]
    =
    \exp\left[
        i\mu k-\frac{\sigma^2 k^2}{2}
    \right],
    \label{eq:g_fourier}
\end{equation}
and define the remainder by $\tilde r(k)=\tilde p(k)-\tilde g(k)$. Equivalently, in real space,
\begin{equation}
    p(x)=g(x)+r(x).
    \label{eq:p_g_r}
\end{equation}
This specific choice of $g$ is not arbitrary. 
For finite-variance summands, repeated convolution flows toward the Gaussian fixed point selected by the CLT. We therefore organize the exact distribution around this fixed point and keep the non-Gaussian part as a residual correction.
Using Eq.~\eqref{eq:p_g_r}, one has $[\tilde p(k)]^N=[\tilde g(k)+\tilde r(k)]^N$. Expanding binomially and transforming back to real space gives the exact hierarchy
\begin{equation}
    P_N(X)
    =
    \sum_{m=0}^{N}
    \binom{N}{m}
    \left(r^{*m}*G_{N-m}\right)(X).    \label{eq:exact_hierarchy}
\end{equation} 
Here $*$ denotes convolution, $r^{*m}$ is the $m$-fold convolution of the remainder $r$ with itself, and $G_{N-m}=g^{*(N-m)}$ is the Gaussian density obtained by convolving $N-m$ copies of $g$. Namely, $G_{m}\left( X\right)=
    \exp\left[
        -\frac{(X-m\mu)^2}{2m\sigma^2}
    \right]/\sqrt{2\pi m\sigma^2}$  with the convention $G_0\left(X\right)=\delta\left(X\right)$. 
Equation~\eqref{eq:exact_hierarchy} is the Gaussian-remainder hierarchy. It is an exact representation of the $N$-fold convolution $P_N=p^{*N}$, organized according to the number of non-Gaussian remainder factors. To see how the different sectors contribute, we write the first terms explicitly:
\begin{equation}
    P_N(X)
    =
    G_N(X)
    +
    N(r*G_{N-1})(X)
    +
    \binom{N}{2}(r^{*2}*G_{N-2})(X)
    +\cdots .
    \label{eq:hierarchy_terms}
\end{equation}

We now use Eq.~\eqref{eq:hierarchy_terms} to identify which sectors of the hierarchy dominate in different parts of the distribution. 
This is the step that turns the exact representation into a useful approximation. 
The two limiting regimes are the Gaussian bulk, where the CLT applies, and the far tails, where a different mechanism is expected to dominate.
We first consider the bulk of the distribution. In the central scaling regime, $X-N\mu=O(\sqrt{N})$, the Fourier representation is controlled by small values of $k$. Since $g(x)$ has the same normalization, mean, and variance as $p(x)$, the remainder $\tilde r(k)=\tilde p(k)-\tilde g(k)$ vanishes through quadratic order at small $k$. Therefore, terms in Eq.~\eqref{eq:exact_hierarchy} containing at least one factor of $r$ are subleading in the central region. The leading contribution is the zeroth sector of the hierarchy, i.e., $P_N(X)\simeq G_N(X)$. 
Thus, the usual Gaussian approximation is the zeroth-order truncation of the Gaussian-remainder hierarchy.


We next consider the opposite limit, the far tail of the distribution.
We assume that the single-summand density $p(x)$ has a tail broader than any exponential,
\begin{equation}
\lim_{|x|\to\infty}
\frac{\exp(-c|x|)}{p(x)}=0,
\qquad\forall c>0 .
\label{eq:broader_than_exp}
\end{equation}
This class includes stretched-exponential densities, $p(x)\sim \exp[-a|x|^\alpha]$ with $0<\alpha<1$, and power-law tails, $p(x)\sim |x|^{-1-\beta}$, restricted here to $\beta>2$ so that the CLT applies.
For a density satisfying Eq.~\eqref{eq:broader_than_exp}, the Gaussian reference density is asymptotically negligible in the far tail, $g(x)/p(x)\to0$ as $|x|\to\infty$, and therefore convolution with a Gaussian background does not change the leading tail. Namely, 
$(p*g)(X)\sim p(X)$ in the $|X|\to\infty$ limit. 
More generally, for any $m\geq 0$, 
\begin{equation}
(p*G_m)(X)\sim p(X),
\qquad |X|\to\infty .
\label{eq:pg_tail}
\end{equation}  
For the densities considered here, we restrict attention to cases in which the BJP applies~\cite{Chistyakov1964,Denisov2008,Foss2013,Vezzani2019SingleBigJump}. Namely, for fixed $j\ge 1$,
\begin{equation}
    p^{*j}(X)\sim j p(X),
    \qquad |X|\to\infty .
    \label{eq:subexp_convolution}
\end{equation}
This expresses that the tail of a sum of $j$ variables is generated, to leading order, by one large summand and $j-1$ typical ones. We now use Eqs.~\eqref{eq:pg_tail} and \eqref{eq:subexp_convolution} to determine the leading tail of the residual terms in the Gaussian-remainder hierarchy. Expanding
\begin{equation}
    r^{*m}
    =
    (p-g)^{*m}
    =
    \sum_{j=0}^{m}
    \binom{m}{j}
    (-1)^{m-j}
    p^{*j}*g^{*(m-j)}
    \label{eq:rm_expand}
\end{equation}
and convolving with $G_{N-m}$ gives terms of the form $p^{*j}*G_{N-j}$. For $j\ge1$, Eqs.~\eqref{eq:pg_tail} and \eqref{eq:subexp_convolution} imply
\[
    (p^{*j}*G_{N-j})(X)\sim j p(X),
    \qquad |X|\to\infty ,
\]
whereas the purely Gaussian term $j=0$ is negligible compared with $p(X)$. Therefore, the coefficient of the leading $p(X)$ contribution in the $m$th residual term is
\begin{equation}
    \sum_{j=1}^{m}
    \binom{m}{j}
    (-1)^{m-j}j
    =
    (-1)^m
    \left.
    x\frac{d}{dx}(1+x)^m
    \right|_{x=-1}
    =
    0,
    \qquad m\ge2 .
    \label{eq:summationzero}
\end{equation}
Thus, the leading one-big-jump contribution cancels in all residual terms with $m\ge2$, and
$(r^{*m}*G_{N-m})(X)=o(p(X))$, for $m\ge2$ .
For fixed $N$, the binomial prefactors in Eq.~\eqref{eq:exact_hierarchy} do not affect this tail comparison. Hence, the only residual term in Eq.~\eqref{eq:exact_hierarchy} that contributes at leading order in the large $|X|$ limit is the $m=1$ term.

Combining the two limiting regimes, the zeroth-order truncation in Eq.~\eqref{eq:exact_hierarchy} gives the leading Gaussian bulk, while the first residual term gives the leading big-jump tail. This motivates the first-order approximation obtained by keeping precisely these two terms:
\begin{equation}
    P_N^{(1)}(X)
    =
    G_N(X)
    +
    N(r*G_{N-1})(X)
    =
    N(p*G_{N-1})(X)
    -
    (N-1)G_N(X).
    \label{eq:P1}
\end{equation}
The first term in Eq.~\eqref{eq:P1} is not merely a tail correction. It is the density obtained by keeping one summand with its exact distribution $p$, while replacing the remaining $N-1$ summands by their Gaussian fixed-point approximation.
Approximation in Eq.~\eqref{eq:P1} requires only a single convolution between the original density and the Gaussian background. 
The first term allows any one of the $N$ summands to be treated exactly, while the remaining $N-1$ summands are represented by the Gaussian fixed point. The subtraction removes the Gaussian overcounting that would otherwise occur in the central region.

The second-order approximation is
\begin{equation}
    P_N^{(2)}(X)
    =
    P_N^{(1)}(X)
    +
    \binom{N}{2}
    \left[
        (p*p*G_{N-2})(X)
        -
        2(p*G_{N-1})(X)
        +
        G_N(X)
    \right].
    \label{eq:P2}
\end{equation}
It keeps two summands explicitly distributed according to $p(x)$, while the remaining $N-2$ summands form the Gaussian background. In the next section we test these closed approximations for explicit broad finite-variance densities and compare them with the exact $N$-fold convolution, the Gaussian approximation, and the big-jump asymptotics.

\begin{figure}[t]
    \centering
    \includegraphics[width=0.7\linewidth]{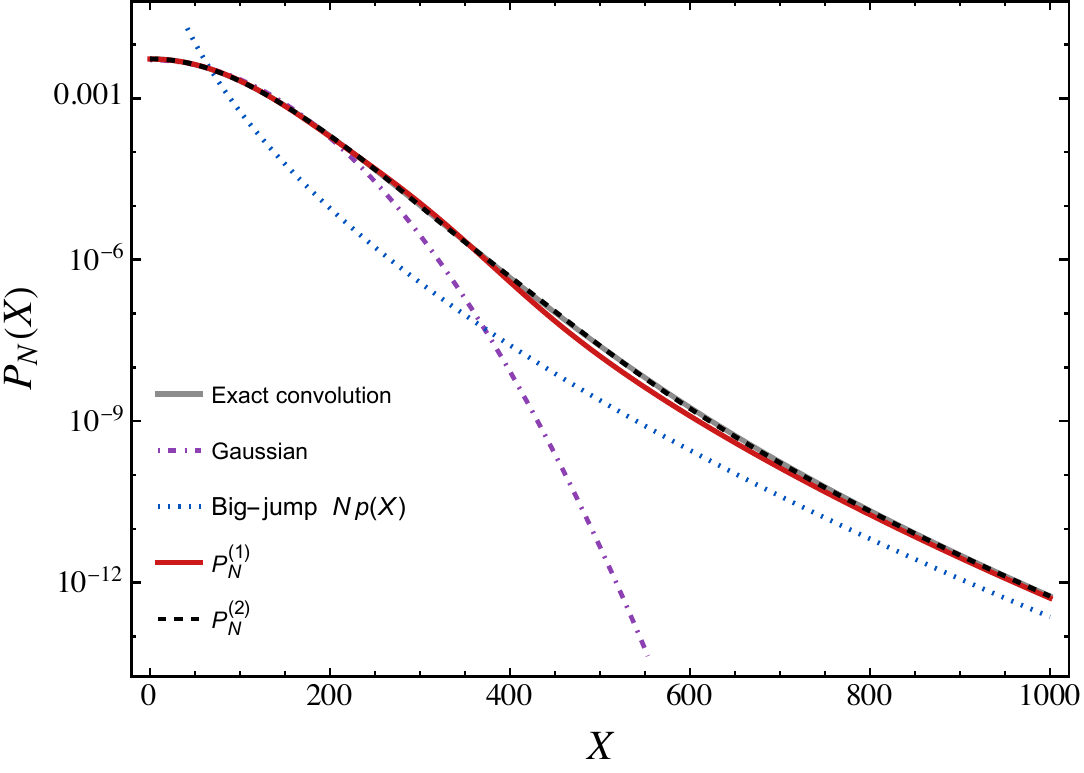}
    \caption{
    Comparison of approximations for the probability density $P_N(X)$ of the sum of $N=50$ independent variables drawn from the symmetric stretched-exponential density
    $p(x)=\frac{1}{4}\exp[-\sqrt{|x|}]$.
    The exact density $P_N(X)=p^{*N}(X)$ is compared with the Gaussian approximation,
    the big-jump asymptotic form $N p(X)$, and the first and second-order
    Gaussian-remainder approximations $P_N^{(1)}(X)$ and $P_N^{(2)}(X)$, as provided by Eq.~\eqref{eq:P1} and Eq.~\eqref{eq:P2}.
    The first-order approximation follows the exact density across the bulk, crossover,
    and tail, while the second-order approximation is nearly indistinguishable from the
    exact result on this scale.
    }
    \label{fig:pdf_comparison_N50}
\end{figure}

\section{Finite-\(N\) Accuracy for a Stretched-Exponential Density}
\label{sec:exampleStretched}

We now test the Gaussian-remainder approximations for an explicit broad finite-variance density. We use the symmetric stretched-exponential law
\begin{equation}
    p(x)=\frac{1}{4}\exp[-\sqrt{|x|}],
    \label{eq:stretched_example}
\end{equation}
which has zero mean and variance $\sigma^2=120$. This example is useful because this $p(x)$ has finite first and second moments, and therefore a central Gaussian region for $P_N(X)$ will appear for large enough $N$.
The subexponential tail of $p(x)$ in Eq.~\eqref{eq:stretched_example} is sufficient for the BJP. 
Between the Gaussian center and the stretched exponential regime at the tails, an anomalous crossover scale separating the two regimes exists and will be discussed in the next section. 
We compare the exact density $P_N(X)=p^{*N}(X)$ with the Gaussian approximation, the big-jump asymptotic form, and the first- and second-order Gaussian-remainder approximations in Eqs.~\eqref{eq:P1} and \eqref{eq:P2}. 
The comparison is performed for several values of $N$, and we use both the density itself and ratios to the exact result to quantify accuracy across the bulk, crossover, and tail regimes.

For $p(x)$ in Eq.~\eqref{eq:stretched_example}, the exact finite-$N$ distribution was evaluated numerically from the Fourier representation in Eq.~\eqref{eq:PN_fourier}. 
The Gaussian approximation is $G_N(X)$, the classical BJP approximation is $Np(X)$, as provided by Eq.~\eqref{eq:subexp_convolution}.
The hierarchy approximations are computed from Eqs.~\eqref{eq:P1} and \eqref{eq:P2}, with the required Gaussian convolutions evaluated numerically. 
Figure~\ref{fig:pdf_comparison_N50} shows the comparison for $N=50$. The Gaussian approximation describes the central region but decays too rapidly and therefore fails in the tail. The classical big-jump approximation converges to the correct far-tail asymptotics, but this convergence is slow, and $Np(X)$ differs substantially from the exact density over much of the range shown. 
In contrast, the first-order hierarchy approximation follows the exact density across the bulk, crossover, and tail, while the second-order approximation is nearly indistinguishable from the exact finite-$N$ distribution on this logarithmic scale.

\begin{figure}[t]
    \centering

    \begin{minipage}{0.48\textwidth}
        \centering
        \includegraphics[width=\textwidth]{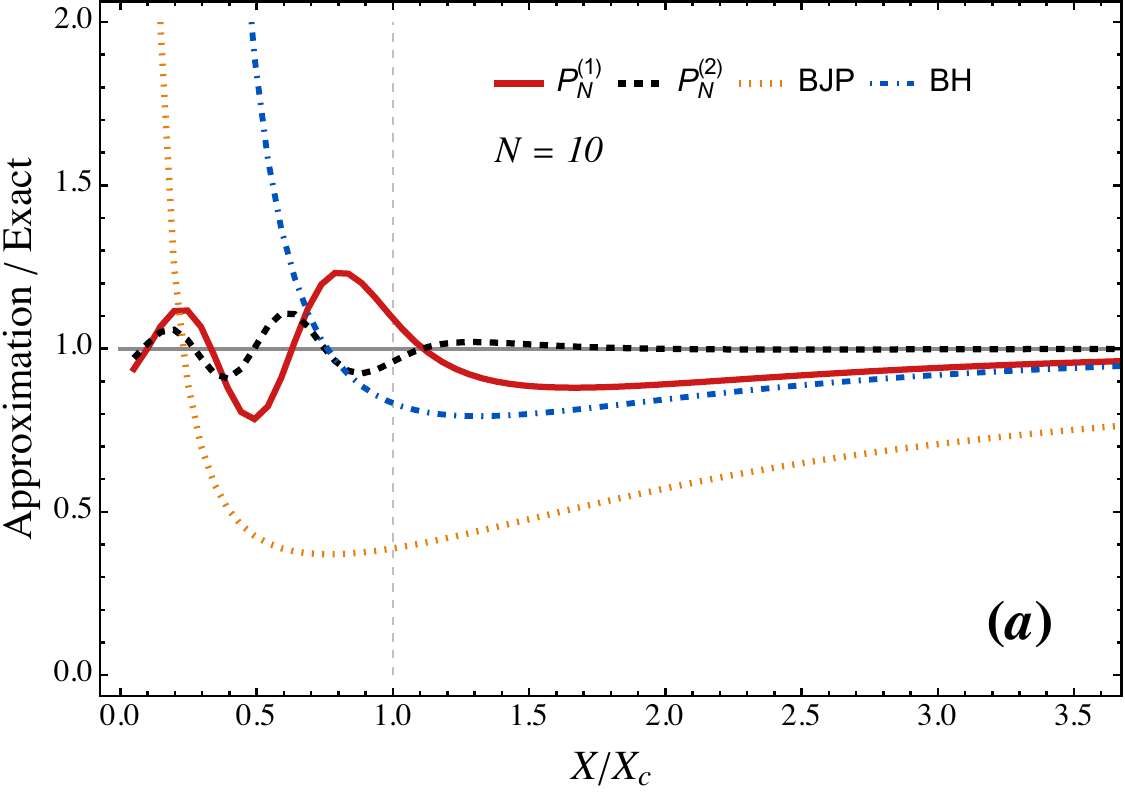}
    \end{minipage}
    \hfill
    \begin{minipage}{0.46\textwidth}
        \centering
        \includegraphics[width=\textwidth]{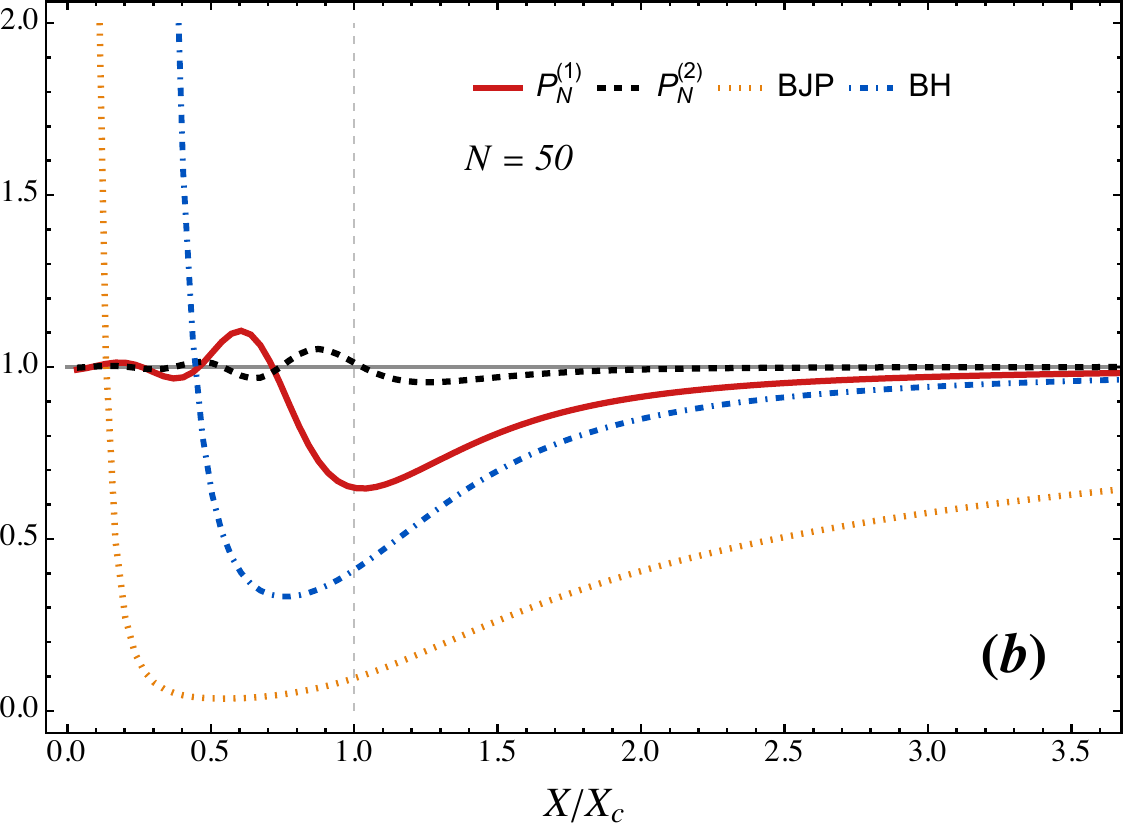}
    \end{minipage}

    \vspace{0.25cm}

    \begin{minipage}{0.48\textwidth}
        \centering
        \includegraphics[width=\textwidth]{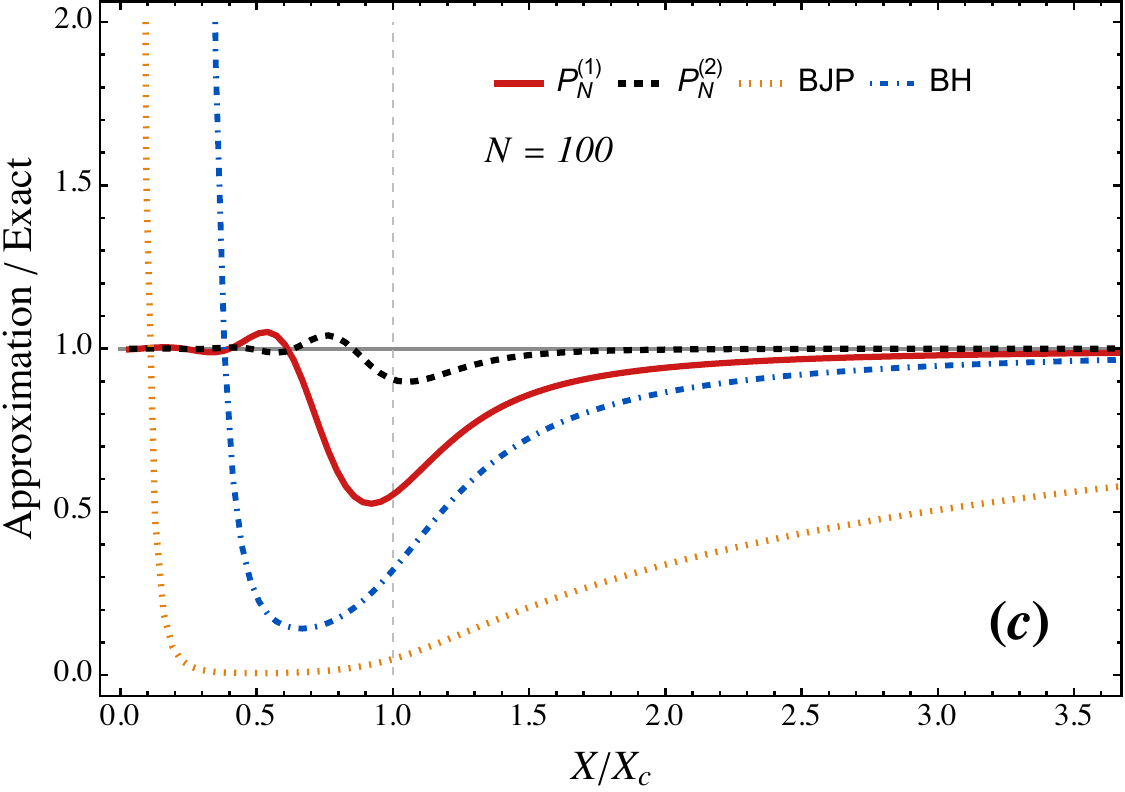}
    \end{minipage}
    \hfill
    \begin{minipage}{0.46\textwidth}
        \centering
        \includegraphics[width=\textwidth]{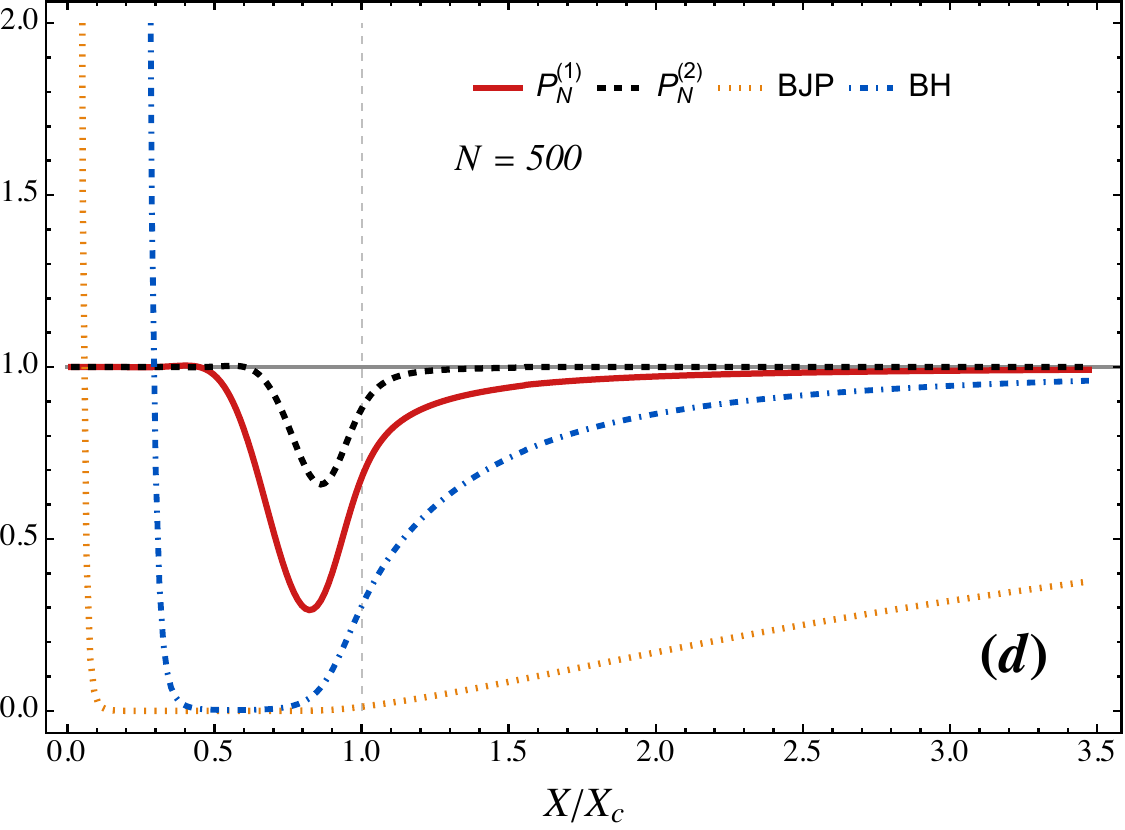}
    \end{minipage}

    \caption{
    Ratio between different approximations and the exact numerical convolution for the stretched-exponential density
    $p(x)=\frac{1}{4}\exp[-\sqrt{|x|}]$.
    The four panels correspond to $N=10,50,100,$ and $500$.
    The horizontal axis is rescaled by the large-deviation crossover scale
    $X_c(N)=\frac{3}{2}(120)^{2/3}N^{2/3}$.
    The plotted quantity is
    $R_{\rm app}(X)=P_N^{\rm app}(X)/P_N^{\rm exact}(X)$, so that the gray horizontal line at $R_{\rm app}=1$ marks exact agreement.
    We compare the first-order Gaussian-remainder approximation $P_N^{(1)}$ in Eq.~\eqref{eq:P1}, the second-order approximation $P_N^{(2)}$ in Eq.~\eqref{eq:P2}, the big-jump asymptotic form $Np(X)$, and the Bassanoni-Hamdi  approximation in Eq.~\eqref{eq:BH_approx}.
    The hierarchy approximations improve substantially over the pure big-jump form over a broad range of $X/X_c$. The accuracy is not uniform: the largest deviations occur near the crossover scale, especially for the first-order approximation as $N$ increases. 
    The second-order approximation reduces these deviations.
    The Bassanoni-Hamdi expression provides a complementary tail-side approximation and improves the pure big-jump estimate, but it is not designed to reproduce the crossover region or the Gaussian bulk.
    }
    \label{fig:ratio_comparison}
\end{figure}

To test the accuracy more directly, we next consider the ratio between each approximation and the exact finite-$N$ density obtained by Fourier inversion. For an approximation $P_N^{\rm app}(X)$ we define
\begin{equation}
    R_{\rm app}(X)
    =
    \frac{P_N^{\rm app}(X)}{P_N(X)} ,
    \label{eq:ratio_definition}
\end{equation}
so that $R_{\rm app}(X)=1$ corresponds to exact agreement. The results are shown in Fig.~\ref{fig:ratio_comparison} for $N=10,50,100,$ and $500$. The horizontal axis is rescaled by the leading crossover scale
\begin{equation}
    X_c(N)=\xi_c N^{2/3},
    \qquad
    \xi_c=\frac{3}{2}(120)^{2/3}.
    \label{eq:Xc_stretched}
\end{equation}
This is the scale at which the anomalous large-deviation theory predicts the crossover between the Gaussian branch and the condensed, single-large-summand branch for the density in Eq.~\eqref{eq:stretched_example}. A detailed discussion of this scale and of the associated transition is given in the next section.
We also compare the developed approximations with the perturbative tail-side approximation of Bassanoni-Hamdi~\cite{BassanoniHamdi2026BeyondBigJump}. For the present density, their Eq.~(24) gives
\begin{equation}
    P_N^{\rm BH}(X)
    =
    Np(X)
    \left(1+\frac{15}{X}\right)^{N-1},
    \qquad X>0 ,
    \label{eq:BH_approx}
\end{equation}
which is constructed as a large-$X$ correction to the classical big-jump form. This approximation provides a useful benchmark from the tail side, but it is not expected to describe the Gaussian bulk or the crossover region.
The ratio plots show that the hierarchy approximations substantially improve over the pure Gaussian and pure big-jump descriptions over a broad range of $X/X_c$. However, the agreement is not uniform. The largest deviations occur in the crossover region, in the vicinity of $X\sim X_c$, where the probability changes from being dominated by collective Gaussian fluctuations to being dominated by a single large summand. This is most clearly seen in panel (d), where for $N=500$ the first-order approximation develops a pronounced dip near the crossover. The second-order approximation reduces this error, but a visible deviation remains. 
The ratio plots also show why the crossover region is the most sensitive part of the distribution: near $X\sim X_c$, the approximation improves when the second hierarchy sector is included, indicating that this region is not captured by a single explicit realization of the broad summand density $p(x)$. 
In this regime, several residual convolutions contribute to the finite-$N$ density, reflecting configurations with more than one atypically large summand on top of the Gaussian background.
These deviations, like in panel (d), are largely hidden on the logarithmic PDF scale of Fig.~\ref{fig:pdf_comparison_N50}, which is why the ratio plot in Fig.~\ref{fig:ratio_comparison} provides a more stringent test of the finite-$N$ accuracy. 


Taken together, Figs.~\ref{fig:pdf_comparison_N50} and \ref{fig:ratio_comparison} provide a proof of concept for the Gaussian-remainder approximation. The first-order truncation provides a good description over a broad range of $X$, while the ratio plots identify the crossover region, $X\sim X_c$, as the regime where higher-hierarchy sectors become relevant. 
These deviations are difficult to resolve on the logarithmic PDF scale of Fig.~\ref{fig:pdf_comparison_N50}, but become apparent in the more sensitive ratio comparison of Fig.~\ref{fig:ratio_comparison}.
In Appendix~\ref{app:power_law_example} we test the $P_N^{(1)}(X)$ and $P_N^{(2)}(X)$ for the case of a finite-variance power-law density. 
The following section focuses on the crossover itself and examines how the same approximations encode the anomalous large-deviation function of $P_N(X)$.

\section{Anomalous rate function and finite-\(N\) corrections}
\label{sec:anomalous_ldf}

We now turn from the PDF accuracy of the hierarchy to its large-deviation structure. 
For sums of independent random variables with light tails, large deviations are usually described by a standard large-deviation principle with speed $N$,
\begin{equation}
    P_N(X\simeq Na)
   \asymp
   \exp[-N I_{\rm st}(a)] ,
   \qquad N\to\infty ,
   \label{eq:standard_ldp}
\end{equation}
where the standard rate function is defined by
\begin{equation}
    I_{\rm st}(a)
    =
    -\lim_{N\to\infty}
    \frac{1}{N}
    \ln P_N(X\simeq Na) .
    \label{eq:standard_rate_definition}
\end{equation}
This is the conventional large-deviation scaling associated with Cramer's theorem~\cite{Touchette2009}. 
For stretched-exponential summand densities, this standard scaling is replaced by an anomalous large-deviation principle. 
This anomalous scaling is a known result for sums with subexponential tails and condensation-type large deviations~\cite{Evans2006CanonicalCondensation,Majumdar2010RealSpaceCondensation,Gradenigo2019RunTumble,Mori2021RTPCondensation,Brosset2020Semiexponential,Smith2025Subleading}.

For concreteness, consider a symmetric density with tail
\begin{equation}
    p(x)\asymp \exp[-K|x|^\alpha],
    \qquad 0<\alpha<1 ,
    \label{eq:stretched_tail_symmetric}
\end{equation}
and finite variance $\sigma^2$. 
The anomalous scaling is
\begin{equation}
    X=\xi N^\gamma,
    \qquad
    \gamma=\frac{1}{2-\alpha},
    \qquad
    \beta=\frac{\alpha}{2-\alpha}.
    \label{eq:anomalous_scaling_general}
\end{equation}
For fixed $\xi$,
\begin{equation}
    P_N(\xi N^\gamma)
    \asymp
    \exp[-N^\beta I(\xi)] ,
    \qquad N\to\infty ,
    \label{eq:anomalous_ldp_general}
\end{equation}
or, in logarithmic form
\begin{equation}
    I(\xi)
    =
    -\lim_{N\to\infty}
    \frac{1}{N^\beta}
    \ln P_N(\xi N^\gamma) .
    \label{eq:anomalous_rate_definition}
\end{equation}
The anomalous rate function is
\begin{equation}
    I(\xi)
    =
    \min_{\eta\ge 0}
    \left[
        K\eta^\alpha
        +
        \frac{(\xi-\eta)^2}{2\sigma^2}
    \right],
    \qquad \xi>0 .
    \label{eq:anomalous_rate_general}
\end{equation}
By symmetry, $I(\xi)=I(|\xi|)$ for $\xi<0$. The nonanalyticity of $I(\xi)$ at the point where the minimizing branch changes is the usual condensation transition of the anomalous large-deviation function.

Our purpose here is not to rederive Eq.~\eqref{eq:anomalous_rate_general} by imposing a separation into different asymptotic regimes. Rather, we show that the same anomalous rate function emerges directly from the Gaussian-remainder hierarchy, already at the first-order truncation.


\begin{figure}[t]
    \centering

    \begin{minipage}{0.49\textwidth}
        \centering
        \includegraphics[width=\textwidth]{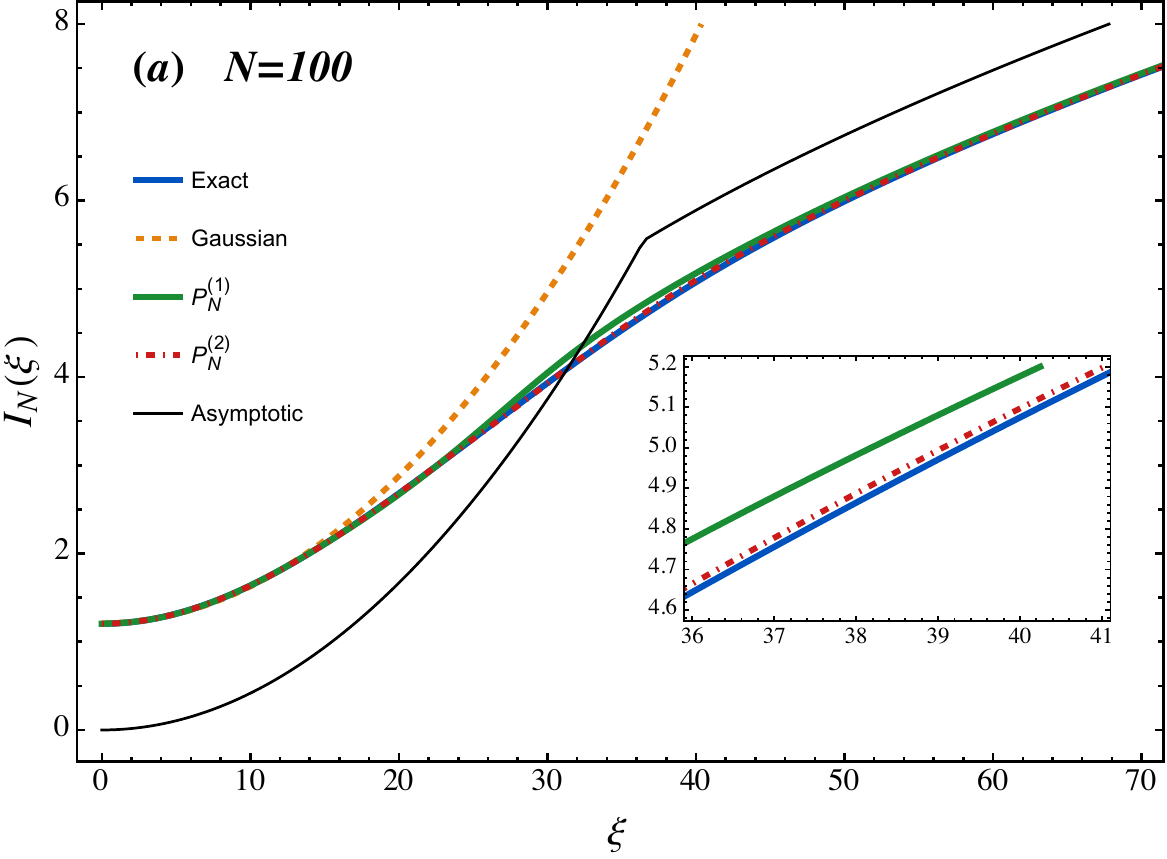}
    \end{minipage}
    \hfill
    \begin{minipage}{0.49\textwidth}
        \centering
        \includegraphics[width=\textwidth]{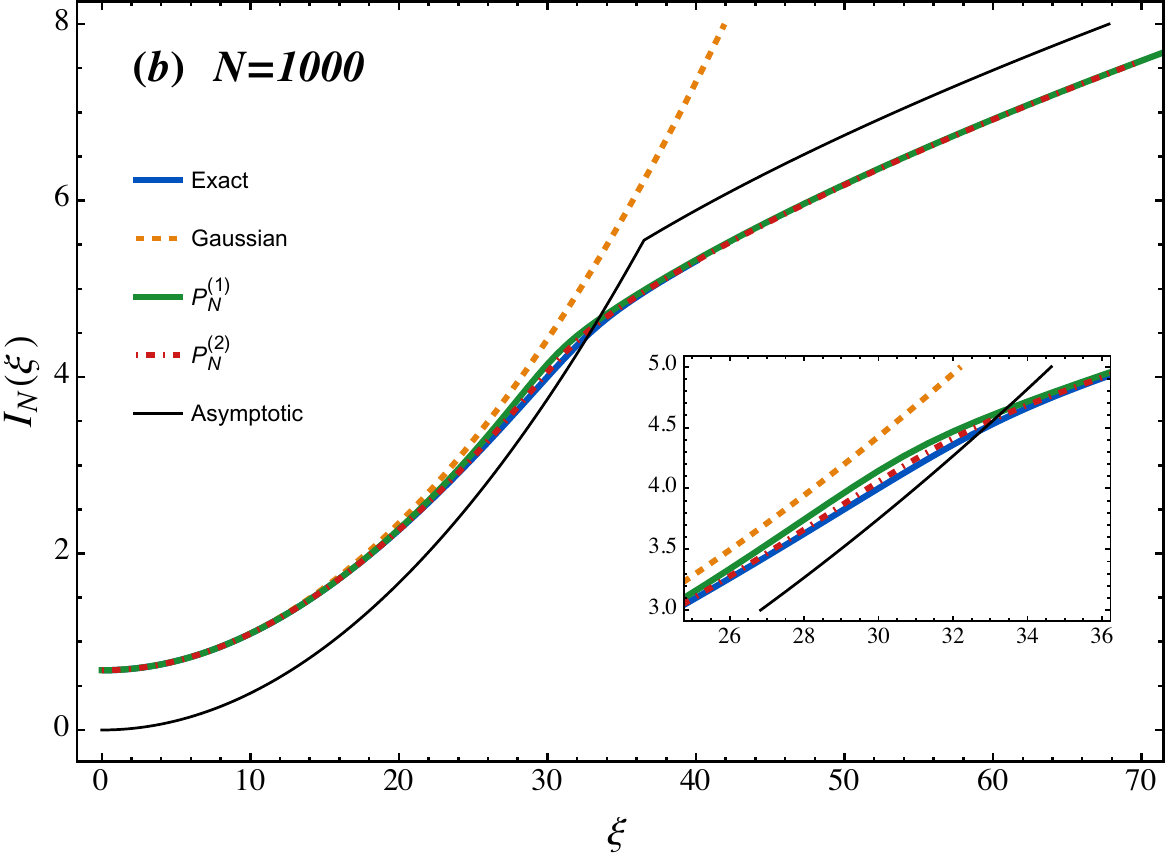}
    \end{minipage}

    \caption{
    Finite-$N$ rate-function comparison for the stretched-exponential density
    $p(x)=\frac{1}{4}\exp[-\sqrt{|x|}]$.
    The plotted quantity is
    $I_N(\xi)=-N^{-1/3}\log P_N(X)$ as a function of
    $\xi=X/N^{2/3}$, corresponding to the anomalous large-deviation scaling for $\alpha=1/2$.
    Panel (a) shows $N=100$, and panel (b) shows $N=1000$.
    The exact numerical convolution is compared with the Gaussian approximation, the first-order hierarchy approximation $P_N^{(1)}$ in Eq.~\eqref{eq:P1}, the second-order approximation $P_N^{(2)}$ in Eq.~\eqref{eq:P2}, and the asymptotic rate function
    $I(\xi)$ in Eq.~\eqref{eq:anomalous_rate_general} with $\sigma^2=120$.
    The hierarchy approximations closely follow the exact finite-$N$ rate function over the plotted range, including the crossover region where the Gaussian approximation fails and the asymptotic rate function displays the limiting cusp structure. 
    The insets in both panels focus on the crossover region. 
    }
    \label{fig:finiteN_rate_function}
\end{figure}

Using Eq.~\eqref{eq:P1}, we write
\begin{equation}
 P_N^{(1)}(X)
 =
N F_1(X)-(N-1)G_N(X),
\qquad
F_1(X):=(p*G_{N-1})(X).
\label{eq:P1_rate}
\end{equation}
We evaluate this expression at $X=\xi N^\gamma$, with $\xi>0$ fixed and with $\gamma,\beta$ given by Eq.~\eqref{eq:anomalous_scaling_general}. Since $\alpha\gamma=\beta$ and $2\gamma-1=\beta$, the convolution $F_1(X)=\int_{-\infty}^{\infty}p(y)G_{N-1}(X-y)\,dy$ gives, at leading exponential order,
\begin{equation}
    F_1(\xi N^\gamma)
    \asymp
    \int_{-\infty}^\infty d\eta\,
    \exp\left\{
    -N^\beta
    \left[
   K\eta^\alpha
   +
  \frac{(\xi-\eta)^2}{2\sigma^2}
    \right]\right\},
\label{eq:F1_anomalous_integral}
\end{equation}
and for large $N$ Laplace's principle yields
\begin{equation}
  -\lim_{N\to\infty}
  \frac{1}{N^\beta}
 \ln F_1(\xi N^\gamma)
 =    I(\xi),
\label{eq:F1_rate_limit}
\end{equation}
where $I(\xi)$ is exactly the same rate function given by Eq.~\eqref{eq:anomalous_rate_general}. The factor $N$ multiplying $F_1$ in Eq.~\eqref{eq:P1_rate} is irrelevant on this scale. 
Thus, the anomalous rate function and the anomalous scaling present themselves in the first-order approximation of the hierarchy. 
The Gaussian term in Eq.~\eqref{eq:P1_rate} is not always negligible. Indeed, this  Gaussian term has rate $\xi^2/(2\sigma^2)$, which is the value of Eq.~\eqref{eq:anomalous_rate_general} at $\eta=0$. If the minimum is attained at $\eta=0$, the two terms in Eq.~\eqref{eq:P1_rate} have the same exponential rate and provide comparable contributions to the final result. The subtraction in Eq.~\eqref{eq:P1_rate} removes only the Gaussian overcounting, as discussed below Eq.~\eqref{eq:P1}, and yields the single-Gaussian asymptotic form. 
If the minimum is attained at $\eta>0$, then $I(\xi)<\xi^2/(2\sigma^2)$, so $NF_1$ is exponentially larger than $G_N$ and the subtraction is negligible. Thus
\begin{equation}
    -\lim_{N\to\infty}
    \frac{1}{N^\beta}
    \ln P_N^{(1)}(\xi N^\gamma)
    =
   I(\xi).
\label{eq:P1_anomalous_rate}
\end{equation}
The first-order hierarchy, therefore, contains the full anomalous rate function.

\begin{figure}[t]
    \centering
    \includegraphics[width=0.7\textwidth]{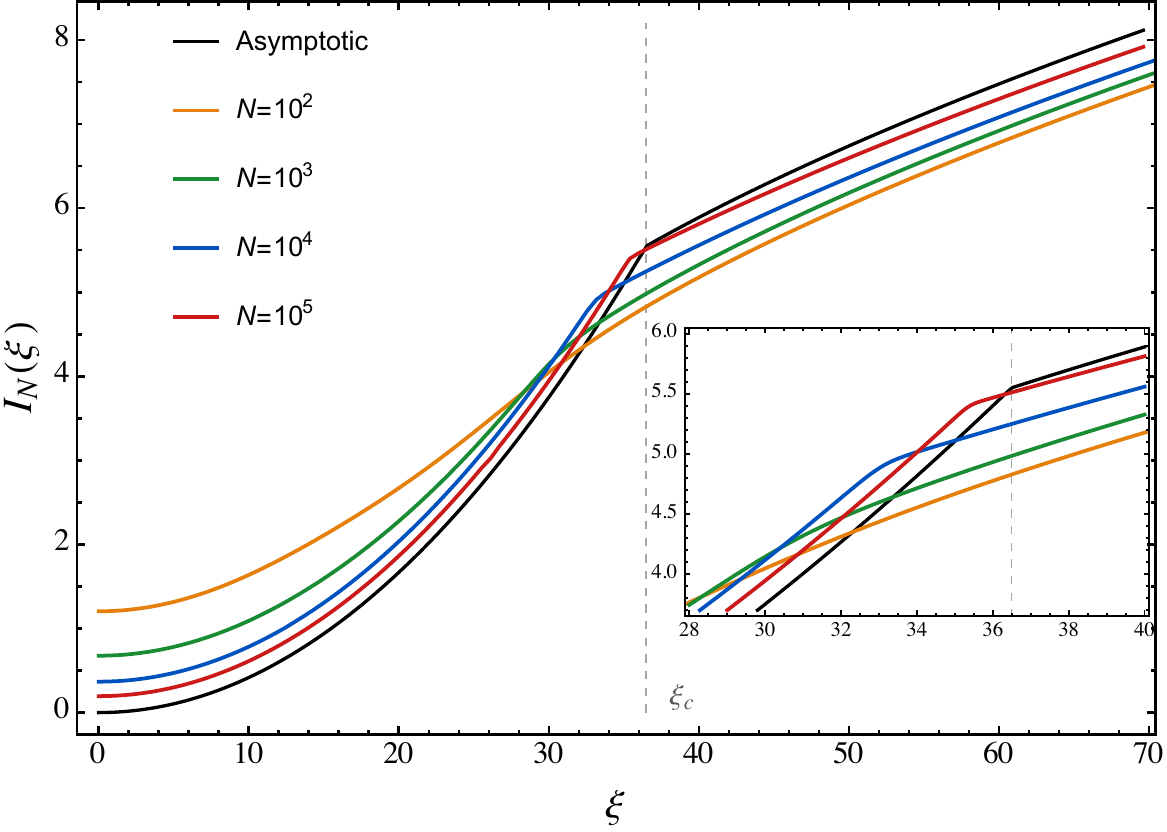}
    \caption{
    Finite-$N$ rate function obtained from the first-order Gaussian-remainder approximation for the stretched-exponential density $p(x)=\frac{1}{4}\exp[-\sqrt{|x|}]$. 
    We plot $I_N^{(1)}(\xi)=-N^{-1/3}\log P_N^{(1)}(\xi N^{2/3})$ for several values of $N$, together with the asymptotic rate function $I(\xi)=\min_{\eta\ge0}\left[\sqrt{\eta}+(\xi-\eta)^2/(2\sigma^2)\right]$, with $\sigma^2=120$.
    The vertical dashed line marks the condensation transition point $\xi_c=(3/2)\sigma^{4/3}$.
    As $N$ increases, the finite-$N$ curves converge toward the asymptotic rate function and develop a sharp form associated with the transition from the Gaussian branch to the condensed, single-large-summand branch.
    The inset focuses on the crossover region and the development of a non-analytical cusp.
    }
    \label{fig:rate_convergence}
\end{figure}

We now consider the higher sectors of the hierarchy in Eq.~\eqref{eq:exact_hierarchy}. 
For fixed $j$, define $F_j(X):=(p^{*j}*G_{N-j})(X)$. Repeating the same scaling argument gives $F_j(\xi N^\gamma)\asymp e^{-N^\beta I_j(\xi)}$ with
\begin{equation}
    I_j(\xi)
    =
\min_{\eta_1,\ldots,\eta_j\ge0}
    \left[
    K\sum_{i=1}^j \eta_i^\alpha
    +
   \frac{
   \left(\xi-\sum_{i=1}^j\eta_i\right)^2
   }{2\sigma^2}
   \right].
   \label{eq:Ij_variational}
\end{equation}
Here, each $\eta_i$ represents the contribution carried by one of the $j$ explicit $p$-distributed summands.
Because $0<\alpha<1$, the stretched-exponential cost is concave, i.e., $\sum_{i=1}^j \eta_i^\alpha \ge \left(\sum_{i=1}^j \eta_i\right)^\alpha$. Therefore, for a fixed total $\eta=\sum_i\eta_i$, the minimum in Eq.~\eqref{eq:Ij_variational} is achieved by placing the anomalous contribution in a single explicit $\eta_i$ and setting all other $\eta$s to zero. Namely, only a single specific $p$-factor in the convolution $F_j$ contributes non-zero displacement in the asymptotic form. 
Hence $I_j(\xi)=I(\xi)$. The corresponding leading contribution has multiplicity $j$, because any one of the $j$ explicit $p$ factors can carry the macroscopic displacement. 
Then we use this asymptotic form of $F_j(X)$ in Eq.~\eqref{eq:rm_expand}, and, according to Eq.~\eqref{eq:summationzero}, it cancels for every $m\geq 2$. Namely, in the large-$N$ limit, there is no asymptotic contribution to the rate function beyond what already appears in the first-order terms. 
 The purely Gaussian contribution is no different: it corresponds to the boundary configuration in which none of the explicit $p$ factors carries a macroscopic displacement. It has no factor $j$, and its coefficient is the binomial sum in Eq.~\eqref{eq:rm_expand} at order zero, which vanishes for all $m\ge1$. 
 Thus, both possible leading contributions of the higher residual sectors are removed by the same subtraction structure that defines the Gaussian-remainder hierarchy.

We established that all the sectors of the Gaussian-remainder hierarchy in Eq.~\eqref{eq:exact_hierarchy} with $m\ge2$ are subleading on the scale $N^\beta$. Configurations with two or more macroscopic summands have a larger stretched-exponential cost, while algebraic prefactors do not affect the rate. Consequently, Eq.~\eqref{eq:anomalous_rate_definition} holds for the first-order correction representation and the
higher hierarchy sectors therefore do not change the anomalous rate function. The role of the higher-order terms is finite-$N$ in nature: they refine the density, especially near the crossover region, where several hierarchy sectors are still numerically visible.

For finite $N$, the limiting anomalous rate function is not necessarily a quantitatively accurate approximation to 
$-N^{-\beta}\log P_N(X)$. 
Figure~\ref{fig:finiteN_rate_function} illustrates this point for the density in Eq.~\eqref{eq:stretched_example}, with $N=100$ in panel (a) and $N=1000$ in panel (b). 
The exact finite-$N$ rate function was obtained from the Fourier representation in Eq.~\eqref{eq:PN_fourier}. 
For this example, $\alpha=1/2$, $K=1$, $\sigma^2=120$, $\gamma=2/3$, and $\beta=1/3$, so that $I_N(\xi)=-N^{-1/3}\log P_N(\xi N^{2/3})$, while the limiting transition point is $\xi_c=\frac{3}{2}(120)^{2/3}$.
The comparison shows that, for the moderate values of $N$ considered here, the asymptotic rate function differs substantially from the finite-$N$ rate function. 
In contrast, the Gaussian-remainder hierarchy provides a much more accurate finite-$N$ description. 
Already the first-order expression $-N^{-\beta}\log P_N^{(1)}(\xi N^\gamma)$ tracks the exact finite-$N$ rate function over most of the range shown. 
The visible deviations are localized near the transition region, where the second-order approximation gives a further improvement. 
Thus, although finite-order truncations still have PDF-level errors, these errors are strongly reduced on the logarithmic rate-function scale, making the hierarchy useful for finite-$N$ large-deviation estimates.    

As shown above, the first-order finite-$N$ rate function,
$-N^{-\beta}\log P_N^{(1)}(\xi N^\gamma)$, converges to the anomalous rate function $I(\xi)$ in the limit $N\to\infty$. 
Figure~\ref{fig:rate_convergence} illustrates this convergence for the density in Eq.~\eqref{eq:stretched_example}. 
The approach is slow: even at $N=10^5$, visible deviations from the limiting rate function remain. 
The inset focuses on the transition region. 
The cusp of the asymptotic rate function at $\xi=\xi_c$ is rounded at finite $N$ and becomes clearly visible only at large $N$, around $N=10^4$ in the present example. 
For low and moderate $N$, the finite-$N$ rate function remains smooth, with no genuine nonanalytic behavior, consistent with the finite-$N$ comparisons in Fig.~\ref{fig:finiteN_rate_function}.


\section{Discussion}
\label{sec:discussion}

We have developed an exact Gaussian-remainder hierarchy for the distribution of the sum of independent, identically distributed random variables with broad, subexponential single-summand densities. 
The construction starts from the decomposition $p=g+r$, where $g$ is a Gaussian with the same normalization, mean, and variance as $p$, and rewrites the full convolution $P_N=p^{*N}$ as the hierarchy in Eq.~\eqref{eq:exact_hierarchy}. 
Truncating this exact expansion gives explicit finite-$N$ approximations. 
In particular, the first-order expression in Eq.~\eqref{eq:P1} involves only a single convolution between the original density and a Gaussian background, together with the subtraction required to preserve the Gaussian sector.

The numerical results show that this simple truncation already provides an accurate approximation to the full density over a broad range of $X$, not only in the asymptotic far tail. 
This was demonstrated for the stretched-exponential density in Eq.~\eqref{eq:stretched_example}, where the first- and second-order approximations were compared with the exact Fourier inversion in Figs.~\ref{fig:pdf_comparison_N50} and \ref{fig:ratio_comparison}. The same construction was also tested for a finite-variance power-law density in Appendix~\ref{app:power_law_example}. 
In both cases, the hierarchy connects the Gaussian central region to the big-jump tail within a single finite-$N$ representation. 
The main deviations from low-order truncations are concentrated near the crossover region, where the ratio plots show that higher-hierarchy sectors become numerically relevant.

The anomalous large-deviation function in Eq.~\eqref{eq:anomalous_rate_general} is not new. 
Its variational structure, which balances a collective Gaussian cost with the cost of a single stretched-exponential summand, belongs to the classical theory of large deviations beyond Cramer's condition, dating back to the work of Nagaev~\cite{Nagaev1969I,Nagaev1969II,Nagaev1979LargeDeviations}. 
It has also appeared more recently in the language of big-jump statistics and condensation transitions~\cite{Brosset2020Semiexponential,Smith2025Subleading}. 
Here we show that the same anomalous rate function emerges directly from the first Gaussian-remainder sector, without imposing a separate decomposition into Gaussian, crossover, and big-jump regimes. 
The higher residual sectors cancel at the leading exponential order and therefore do not modify the limiting rate function.

The same hierarchy also provides practical approximations to finite-$N$ rate functions. 
As shown in Fig.~\ref{fig:finiteN_rate_function}, the limiting anomalous rate function can differ substantially from the finite-$N$ rate function at moderate values of $N$. 
The Gaussian-remainder approximations, however, track the finite-$N$ rate function much more closely on the logarithmic scale. 
The first-order term contains the limiting anomalous rate function, while higher orders refine the density and improve the crossover region. 
Thus, the hierarchy separates the asymptotic exponential structure from the finite-$N$ corrections that remain visible before the large-$N$ limit is reached.

The present construction assumes that the single-summand density has finite variance, so that the Gaussian fixed point is the natural background. 
This is the appropriate setting for the stretched-exponential and finite-variance power-law examples considered here. 
A natural extension is to distributions for which the first or second moment diverge. 
In that case, the Gaussian background should be replaced by the relevant stable law, and the corresponding remainder hierarchy may provide finite-$N$ approximations for sums attracted to L\'{e}vy-stable fixed points. 
It would also be interesting to generalize the method beyond independent sums, for example, to correlated stochastic processes, coupled continuous-time random walks, run-and-tumble dynamics and generalized L\'{e}vy walks, where big-jump ideas have already proved useful in describing rare events~\cite{Vezzani2019SingleBigJump,Vezzani2020LevyWalksBigJump}. 
In such systems, finite-time rare-event distributions often reflect a competition between collective fluctuations and a single dominant event. 
Identifying this event can be physically important, as in recent work where rare transport events lead to anomalous mobility enhancement~\cite{shafir2024disorder}. 
This perspective may also be useful for transport problems in which rare displacement tails are not produced by a single event. 
For example, universal Laplace tails arise from the cumulative contribution of many displacements and fluctuations in the number of renewals, rather than from a single exceptionally large jump~\cite{BarkaiBurov2020PRL,wang2020large}. 
Related studies have shown that, when the jump-length distribution crosses into the subexponential class, this collective mechanism gives way to single-event statistics described by the big-jump principle~\cite{singh2023universal}. 
The hierarchy developed here provides a finite-$N$ framework for such questions: it keeps the collective background explicitly, while successive residual sectors identify the role of one or more atypical summands. 
Extending this construction to continuous-time random walks and other renewal processes may therefore help describe the finite-time crossover between Laplace-type collective tails and single-event-dominated rare events.

\begin{acknowledgments}
This work was supported by the Israel Science Foundation, Grant No.~3791/25.
\end{acknowledgments}

\clearpage
\appendix
\setcounter{equation}{0}
\renewcommand{\theequation}{A\arabic{equation}}
\setcounter{figure}{0}

\renewcommand{\thefigure}{A\arabic{figure}}

\section{Power-law example}
\label{app:power_law_example}

\begin{figure}[b]
    \centering
    \includegraphics[width=0.7\linewidth]{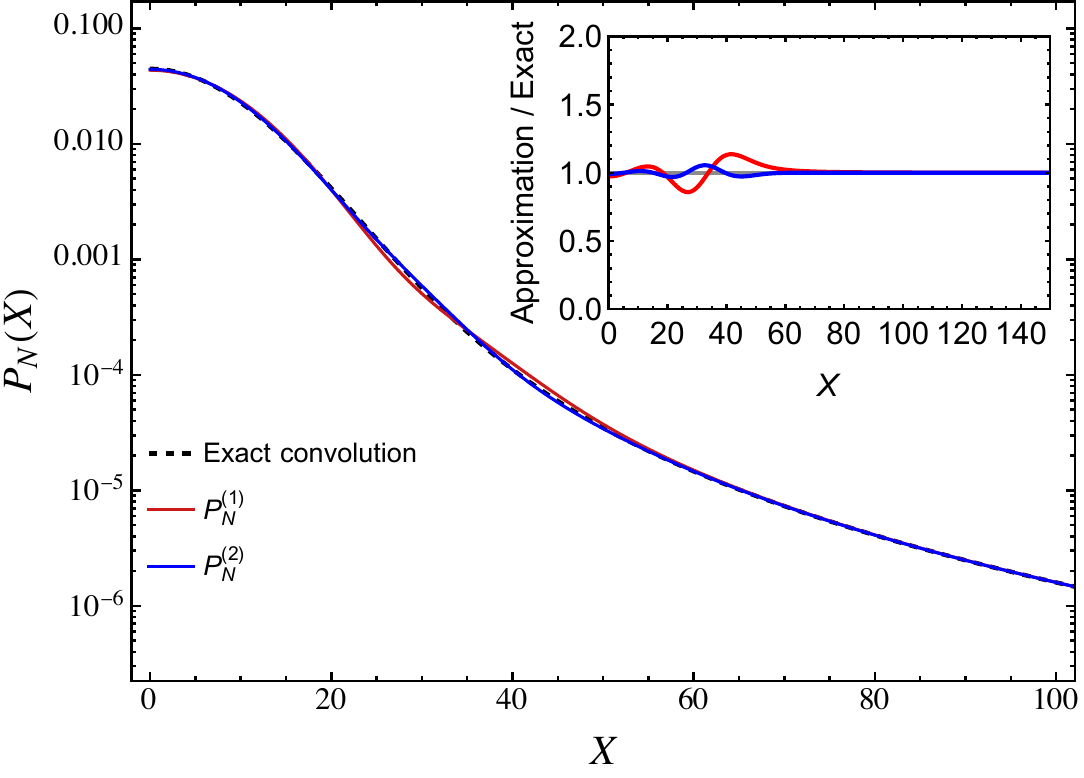}
    \caption{
    Finite-$N$ comparison for the power-law single-summand density
    $p(x)=\frac{\beta}{2}(1+|x|)^{-1-\beta}$ with $\beta=3$ and $N=100$.
    The black dashed curve is the exact numerical convolution $P_N(X)$, the red curve is the first-order hierarchy approximation $P_N^{(1)}(X)$ from Eq.~\eqref{eq:P1}, and the blue curve is the second-order approximation $P_N^{(2)}(X)$ from Eq.~\eqref{eq:P2}. The inset shows the ratios $P_N^{(1)}(X)/P_N(X)$ and $P_N^{(2)}(X)/P_N(X)$. The agreement confirms that the Gaussian-remainder hierarchy is not specific to stretched-exponential tails; for finite-variance power-law tails it also captures the Gaussian center, the crossover region, and the algebraic big-jump tail.
    }
    \label{fig:power_law_appendix}
\end{figure}

In the main text we focused on stretched-exponential densities, for which the crossover between the Gaussian bulk and the big-jump tail is associated with an anomalous large-deviation scale. The Gaussian-remainder hierarchy itself, however, is not restricted to stretched-exponential tails. In this appendix we give a simple check for a finite-variance power-law density.

We consider the symmetric density
\begin{equation}
    p(x)
    =
    \frac{\beta}{2}
    \left(1+|x|\right)^{-1-\beta},
    \qquad
    \beta=3 .
    \label{eq:power_law_example_density}
\end{equation}
For $\beta>2$ the variance is finite, and therefore the central part of the sum is still governed by the Gaussian fixed point of the central limit theorem. At the same time, the density has a subexponential tail, so the far tail is governed by the big-jump principle. This example therefore tests whether the hierarchy can connect the Gaussian center and the algebraic big-jump tail in a case different from the stretched-exponential example used in the main text.

Figure~\ref{fig:power_law_appendix} compares the exact numerical convolution with the first- and second-order hierarchy approximations, Eqs.~\eqref{eq:P1} and \eqref{eq:P2}, for $N=100$. The first-order approximation already follows the exact distribution over the full range shown, including both the central region and the power-law tail. The second-order approximation further reduces the residual finite-$N$ error in the crossover region. The inset shows the corresponding ratios to the exact convolution, making visible that the remaining deviations are localized in the intermediate region where neither the pure Gaussian approximation nor the asymptotic big-jump expression alone is expected to be uniformly accurate.
\vspace{1.2cm}

\bibliography{bjpreferences.bib}

\end{document}